\documentstyle[prb,aps,eqsecnum]{revtex}
\begin{document}
\twocolumn
\baselineskip 0.6cm

\title{The magnetoresistance in GdNi: Magnetic-polaronic-like effect near
Curie temperature and low temperature sign reversal }
\author{R. Mallik, E.V. Sampathkumaran, P.L. Paulose and V. Nagarajan }
\address{ Tata  Institute  of  Fundamental  Research,  Homi   Bhabha
Road,   Colaba, Mumbai-400005, INDIA }
\maketitle

\noindent{\underline{\bf Abstract}}

\noindent The results of magnetoresistance  ($\Delta \rho /\rho$ )
measurements in GdNi, in the temperature range  4.2 to 300 K  are
reported.  The  sign  of $\Delta \rho /\rho $ above the Curie temperature
($T_{C}$= 70 K) is negative and its magnitude in a magnetic field of 80
kOe grows with decreasing temperature below  150 K with a peak value of
about -20$\%$ at $T_{C}$. These features, qualitatively resembling those
in giant magnetoresistance systems (manganates), are attributed  to the
formation of some kind of magnetic polarons induced by Gd.  The
magnetoresistance changes sign below  12 K which  is  attributed  to
subtle  band structure effects.

\vskip 1cm
\noindent{\underline{\bf Introduction}}

\noindent The observation \cite{1} of giant and colossal magnetoresistance
in perovskites, La$_{1-x}$A$_{x}$MnO$_{3}$ (A= Ca, Sr, Ba),  near  the
semiconductor-metal transition temperature has generated much interest in
this family of compounds. There have been lively discussions in recent
literature on the relative importance of  double exchange
\cite{2}   and  Jahn-Teller  electron-phonon coupling mechanisms \cite{3}
for the increase  in  resistivity  drop  with  the application of magnetic
field  (H)  at  the  Curie  temperature ($T_{C}$).  Very recently,
\cite{4}  large magnetoresistance in  amorphous  magnetic  rare earth  (R)
alloys due to the formation of a dense concentration  of magnetic polarons
has been reported. In this context, we considered it worthwhile to  probe
the magnetoresistance $(\Delta \rho /\rho )$ behavior of {\it normal}
(where the double exchange  or  Jahn-Teller mechanisms  are  not expected
to  operate)   ferromagnetic intermetallic compounds near $T_{C}$.  At
this point, it may be stated that we have recently reported interesting
$\Delta \rho /\rho $ behavior  in several  rare-earth intermetallic
compounds across the   N\'eel temperature \cite{5}  as well as in
paramagnetic alloys\cite{6} (in addition to ferromagnetic and
antiferromagnetic Ce systems). As a  continuation  of  our programme in
this direction, we  report  here  the   results   of our magnetoresistance
study on a ferromagnetic  Gd  alloy, viz., GdNi ($T_{C}$= 70
K)\cite{7,8,9,10,11,12}.  The magnetoresistance above $T_{C}$ exhibits a
temperature dependence similar to that noted in La manganates, as if there
is a  formation of magnetic-polarons  (presumably  induced in the  Ni 3d
orbital, which is otherwise nonmagnetic) by  Gd 4f magnetism.  In
addition, we also note an interesting feature at low temperatures, viz., a
positive contribution dominating below 12 K. The extension of present
studies to a few  La  and Y substituted  alloys  of  GdNi also enabled us
to identify the influence of Gd  sublattice dilution, crystal structure
and the chemical pressure effects on the observed features.

\noindent{\underline{\bf Results and Discussion}}

\noindent The  series  RNi   presents   an   interesting  scenario both
from   the crystal structure and magnetism point of view. While La, Ce, Pr
, Nd, Gd and Tb compounds of this  series  have  been  reported  to
crystallize in the CrB-type orthorhombic structure, the heavy rare earth
members (Dy to Tm  as well as  Y)  form in the FeB-type orthorhombic
structure.
\cite{7,8,9,10} We have recently shown \cite{11,12}  that the
exchange interaction is stronger in the FeB  structure  than  in  the CrB
structure. While it is known that Ni does not possess any magnetic moment
in light  R  members of the series RNi, an additional moment beyond  the
value expected for trivalent  Gd was found in GdNi, in the para  as  well
as  in  the ferromagnetic states, induced by  Gd  presumably in Ni 3d
band. On the basis of the magnetization  behavior,   we   have   proposed
\cite{11,12} the coexistence  of  localised  magnetism  (from  Gd 4f
orbital)  and induced itinerant magnetism in the  Ni 3d  band in the
magnetically ordered state.

\par
The alloys chosen for the present study,  GdNi, Gd$_{0.75}$R$_{0.25}$Ni
and Gd$_{0.5}$R$_{0.5}$Ni (R= Y  and  La), were the same as those
investigated previously.\cite{12} The polycrystalline samples were
rectangular in shape (about 2mm $\times$ 2mm cross section; voltage leads
separation about 2-4 mm).  The magnetoresistance, $\Delta
\rho /\rho  [= \{\rho (H)-\rho (0)\}/\rho (0)]$,  data were obtained as a
function of temperature in the range 4.2 - 300 K  in the presence  of an
external magnetic field (H) of 80 kOe and also as  a  function  of H   at
select temperatures in the longitudinal mode. We have employed a constant
current typically of 10 mA and we notice that the results are reproducible
with higher excitation currents (e.g., 50 mA). The voltages were measured
to an accuracy of 10 nV. A conducting silver epoxy was used to make the
electrical contacts of the leads with the samples.  The values of the
resistivity for all the alloys at 300 K are the same, typically 200
$\mu\Omega$ cm, within the estimated error of about 40\%. This error
arises due to uncertainties in measurement of the separation between
voltage leads, caused by spread of the silver paste. Due to this reason,
only the resistance data is presented in the figures.

\par
The temperature dependence of $\rho $ in an applied field of 80  kOe as
well as in zero field, along with the temperature dependence of
$\Delta\rho /\rho $  obtained  from this data, are shown in Fig. 1 for
GdNi. Since the value of $\Delta \rho /\rho $ above  200 K  is negligible,
the data are shown only below this temperature. The value of  $\Delta
\rho /\rho $, though  small,  is  negative  around  140 K. Its
magnitude progressively grows as $T_{C}$ is approached from higher
temperatures, exhibiting  a maximum (about $-20\%)$ at $T_{C}$. This is
followed by a continuous fall in magnitude as the temperature is lowered
further. The temperature dependence of $\Delta \rho /\rho $ observed here
mimics the behavior reported in  La  manganates in bulk form. \cite{13}
Though the magnitude of $\Delta \rho /\rho $ at $T_{C}$ is not as large as
in La manganates, it is  still  large  for a metal, considering that the
Lorentz force on the conduction electrons is generally known to cause a
positive value of $\Delta
\rho /\rho $ of less than a  few  percent.  While the behavior in La
manganates and the amorphous magnetic rare earth  silicon alloys
\cite{1,2,3,4}  is related to the metal-insulator transition at $T_{C}$,
that  in  GdNi cannot be due to such a factor, as it remains metallic in
the  entire temperature range of investigation. As in the case of
GdNi$_{2}$Si$_{2}$ (Ref. 5), the negative value suggests that there are
spin fluctuations in the  Ni 3d band induced by  Gd 4f  magnetic moment,
that are suppressed by the application of the magnetic field. The measured
effective moment is noticeably larger than that expected for trivalent  Gd
ion and we have proposed
\cite{12}   that  the  induced moment (in other words, some kind of
polaron) above $T_{C}$ is of a localised type. The localised magnetic
moment proposed to be from Ni 3d orbital was deduced \cite{12}   to be
close to $3 \mu _{B}$. The increasing magnitude of $\Delta \rho /\rho $
with a {\it negative sign}  as $T_{C}$ is approached from higher
temperatures indicates a  gradual increase in the  suppression  of  spin
fluctuations (from the polarons) by  the applied field.  To understand the
origin of polaron formation, we propose the existence  of short range (of
the order of mean free path of the  conduction  electrons) magnetic
ordering in  Gd sublattice.  The  heat  capacity  results \cite{8,12}
provide evidence for  this  proposal  in  the  sense  that  the  magnetic
contribution to heat capacity  has been shown to exhibit \cite{8,12} a
continuous upturn  over  a wide temperature range before long range
magnetic order sets in. The  Ni 3d orbital is more prone to  polarisation
by  this short range  order, as demonstrated\cite{5} by a comparative
study of GdCu$_{2}$Si$_{2}$ and GdNi$_{2}$Si$_{2}$;  the short range order
does not seem to polarise the  Cu 3d  band as  the magnetoresistance
before long range magnetic order  sets  in is positive in this Cu
alloy.  Apparently,  the  5d   and  6s electrons  do   not contribute
significantly to the polaron formation, as otherwise one should  have seen
negative $\Delta \rho /\rho $ above $T_{N}$ in GdCu$_{2}$Si$_{2}$ as well.
Thus,  there is some justification for our  proposal that the polarons are
constituted mainly by Ni 3d  orbital.  The present results also emphasize
that the short  range  magnetic  order (from Gd),  not necessarily long
range order, is  sufficient  to cause this polaronic-like effect. Well
below $T_{C}$, the polarization cloud apparently becomes  itinerant,
coupling ferromagnetically with the  Gd 4f moment;\cite{12}  as a result
the (localised) polaronic contribution is turned off below $T_{C}$,  as
proposed  by Millis et al \cite{3}  for La$_{1-x}$Sr$_{x}$MnO$_{3}$.
Therefore, below $T_{C}$, this prominent source for negative  contribution
to $\Delta \rho /\rho $ diminishes.
\par
An interesting feature observed here is the change in  the  sign  of
$\Delta \rho /\rho $ below  12 K (see Fig. 1). To get a better view of the
sign and non-monotonic variation of the magnitude of $\Delta
\rho /\rho $  with  temperature,  we present $\Delta \rho /\rho $ as
a function of field at various temperatures in Fig. 2.  In a field of 60
kOe, the values of $\Delta \rho /\rho $ are about $-7\%, -15\%, -7\%$,
zero, and $12\%$  at  95, 50, 20, 12 and  4.2 K  respectively. These
values are well outside the experimental error  of $\pm$1\% (obvious from
the scatter in the values as a function of field).  For instance, for GdNi,
at 4.2 K,  the observed voltage difference for H= 0 and H= 60 kOe is
typically 1$\mu$V, which is far above the sensitivity of the
nanovoltmeter; thus the magnetoresistance values are reliable within the
experimental error. In the paramagnetic state, for instance, at 95 K, it
is clearly negative (presumably varying  as $H^{2}$) as expected for spin
fluctuating  systems.  The positive contribution however appears at 4.2 K.
The positive peak in the low field data for 20 K is of no significance
considering an error of $\pm$1\%.  The observation of  positive sign of
magnetoresitance  at 4.2 K with a large magnitude at higher fields is
quite interesting and reveals the existence of two competing contributions
to $\Delta \rho /\rho $.  These two contributions appear to be  of nearly
equal magnitude at 12 K, as the values are close to zero at all fields at
this temperature. The competition between these interactions is visible as
a shoulder in $\Delta \rho /\rho $ around 25 K (Fig.  1).  The sign
crossover and large magnitude have been found by us at low temperatures
well below $T_{C}$ in other ferromagnetic systems (from Mn sublattice)
like SmMn$_{2}$Ge$_{2}$ (Ref. 14) and LaMn$_{2}$Ge$_{2}$ (Ref. 15). It is
also possible that the negative contribution from domain walls is frozen
around 25 K, as a result of which the  positive contribution from
conduction electrons dominates.  It  is of interest to explore whether the
application of the field causes an unusual effect  of inducing gaps in
some portions of  the Fermi surface, resulting in changes in the band
structure. The data on  La and  Y substituted alloys presented below
appear to favor this view, though more work is required to completely
understand this interesting aspect of $\Delta
\rho /\rho $.

\par
The results on the  La   and  Y   substituted  alloys,
Gd$_{0.75}$La$_{0.25}$Ni ($T_{C}$= 52 K), Gd$_{0.5}$La$_{0.5}$Ni ($T_{C}$=
35 K), Gd$_{0.75}$Y$_{0.25}$Ni ($T_{C}$= 55 K) and Gd$_{0.5}$Y$_{0.5}$Ni
($T_{C}$= 65 K) (Refs. 11, 12), are shown  in Fig. 3.  While the CrB-type
structure  is  maintained  in  the  former three alloys, the last one
crystallizes  in  the  FeB  type  structure.
\cite{10,11,12}  While  La expands the lattice,  Y  compresses the lattice.
Thus  the  data  on these alloys is useful to identify the influence of
crystal  structure and chemical pressure effects on the observed features.
The  common feature  in all these alloys is that $\Delta \rho /\rho $
peaks at $T_{C}$. The magnitude decreases with  the decrease in the  Gd
concentration, rendering support to  our  idea  that  the observed $\Delta
\rho /\rho $  effect  is  Gd   induced  (polaronic-like)  effect;  apparently
 La  substitution is more effective compared to  Y  in depressing the
magnitude  of $\Delta \rho /\rho $. The temperature below which the
polaronic-like effect dominates appears  to  be about $2T_{C}$, as in the
case of GdNi.  There are some qualitative differences in the low
temperature  range ($<$12 K).  It  is  apparent  that,  in  all  these
substituted alloys, the value of $\Delta \rho /\rho $ remains negative
even  below  12 K.  We have also slightly changed the  Gd/Ni  composition
by about $5\%$  and  the $\Delta \rho /\rho $ remains negative (not shown
here) down to 4.2 K. These findings suggest  that the low temperature
positive $\Delta \rho /\rho $ behavior in  stoichiometric  GdNi  may  be
related to subtle band structure effects at the Fermi level $(E_{F})$,
which  get modified by the shift of $E_{F}$ by the unit cell volume
changes.

\noindent{\underline{\bf Conclusion}}

\noindent In summary, we have  reported  here  a  gradual  increase  of
negative magnetoresistance below twice of Curie temperature till the onset
of long range magnetic ordering in GdNi. We attribute this to some kind of
magnetic polaronic contribution. Thus, this article reports polaronic-like
effects in the magnetoresistance in a  ferromagnetic  rare earth
$\it{intermetallic}$ compound, that is, a material in which neither
double-exchange nor Jahn-Teller  mechanisms are expected to be operative.
The findings suggest that the contribution of spin fluctuations from
magnetic polarons may have to be considered in the giant magnetoresistance
systems as well. In addition, for $T<<T_{C}$, positive  values are  noted
in  this compound.

\vskip 0.5cm
\noindent{\underline{\bf Figure Captions}}

\begin{figure}  
\caption{Electrical resistance as a function of temperature for GdNi in
the presence and in the absence of a magnetic field of 80 kOe.  The
magnetoresistance ($\Delta \rho /\rho $)  obtained  from this data are
also shown.}
\end{figure}

\begin{figure}
\caption{The  magnetoresistance ($\Delta \rho /\rho $) as  a  function  of
magnetic   field at   various temperatures for GdNi. The lines (except for
12 K) through the data points  serve as a guide to the eye. The
experimental error in the values at low temperatures ($\le$ 20 K) is larger
($\pm$1\%) than at higher temperatures due to smaller values of resistance.
Though the low field peak for 20 K is not genuine due to this reason, we
have drawn a line through the data points for this temperature as well for
the sake of clarity.}
\end{figure}

\begin{figure}
\caption{Electrical resistance as a function of temperature for
Gd$_{1-x}$R$_{x}$Ni (R= Y and La; $x$= 0.25 and 0.5) in the presence and
in the absence of a magnetic field of  80  kOe.  The magnetoresistance
($\Delta \rho /\rho $) obtained  from this data are also shown.}
\end{figure}

\end{document}